# PHASE GROUPING OF LARMOR ELECTRONS BY A SYNCHRONOUS WAVE IN CONTROLLED MAGNETRONS*

G. Kazakevich[#], R. Johnson, Muons, Inc, Batavia, IL 60510, USA
V. Lebedev, V. Yakovlev, Fermilab, Batavia, IL 60510, USA

*Abstract*
A simplified analytical model based on the charge drift approximation has been developed. It considers the resonant interaction of the synchronous wave with the flow of Larmor electrons in a magnetron. The model predicts stable coherent generation of the tube above and below the threshold of self-excitation. This occurs if the magnetron is driven by a sufficient resonant injected signal (up to -10 dB). The model substantiates precise stability, high efficiency and low noise at the range of the magnetron power control over 10 dB by variation of the magnetron current. The model and the verifying experiments with 2.45 GHz, 1 kW magnetrons are discussed.

## INTRODUCTION

Magnetrons are used in normal conducting accelerators as efficient and inexpensive RF generators. For superconducting accelerators, the magnetron generators must provide the fast phase and power control necessary to stabilize the accelerating voltage in the Superconducting RF (SRF) cavities [1]. Methods for fast control of the phase and power in magnetrons driven by resonant injected signals have been recently developed [2-4]. The methods allow for the precise stabilization of the phase and amplitude of the accelerating voltages in multi-cavity superconducting accelerators. A "Reflecting amplifier" model presently used for evaluating the properties of RF-driven magnetrons [5], describes the forced oscillation in the vicinity of the resonance and do not consider the impact of the driving signal on the magnetron stability, efficiency, noise. The developed model considers the energy exchange of the electron flow with the self-consistent field of the synchronous wave as the basing principle of coherent generation of magnetrons. The model allows for the evaluation of the necessary and sufficient conditions for the coherent generation of the magnetron. It predicts the stable operation of a magnetron below the Hartree voltage when the tube is driven by a resonant signal of sufficient amplitude. It also substantiates the impact of the driving signal on the magnetron efficiency and noise.

## CONDITIONS FOR COHERENT GENEERATION OF MAGNETRONS

The RF energy in a magnetron RF system is determined by the static electric field [6]. The injected resonant signal increases the RF energy in accordance with the energy conservation law. Therefore, this is equivalent to an increase of the magnetron voltage. This allows stable operation of the RF driven tube at voltages below the Hartree voltage [4]. In this case the magnetron current can be significantly less than the minimum current allowed in free run operation. This provides an extended range of power control (*R*) by variation of the magnetron current in the extended range. Considering the energy balance in the RF system of the magnetron, one can estimate the equivalent increase of the magnetron voltage due to the resonant injected signal and the extended range of power control [4]. For magnetrons at injected signal power of -10 dB), one obtains $R \sim 10$ dB.

## INTERACTION OF THE SYNCHRONOUS WAVE WITH THE DRIFTING CHARGE

We consider the simple analytical model based on the charge drift approximation [6], for conventional CW N-cavity magnetrons driven by a resonant RF signal. We discuss the magnetron with a constant uniform magnetic field $H$, above the critical magnetic field. The magnetron operates in the $\pi$-mode, i.e. with the RF electric field shifted by $\pi$ in neighbouring cavity gaps.

The magnetron operates at the cyclotron frequency $\omega$, with a well matched load. In the drift approximation we consider the motion of charge in the center of the Larmor orbit with radius $r_L$, when the Larmor motion of the electron itself is averaged over the frequency $\omega$ [6]. We neglect the impact of space charge since we consider CW tubes operating at rather low currents. In this approach we neglect the azimuthal non-uniformity of the static electric field. The drift of charge in the center of the Larmor orbit with azimuthal angular velocity $\Omega$ is determined by the superposition of the static electric field described by the potential $\Phi^0$ and the RF field of the synchronous wave that is induced by the magnetron current and the injected resonant signal. The RF field can be determined by a scalar potential $\Phi$. Thus, the drift of the charge is described in the polar frame by the following system of equations [7]:

$$\begin{cases} \dot{r} = -\dfrac{c}{Hr}\dfrac{\partial}{\partial \varphi}(\Phi^0 + \Phi), \\ \dot{\varphi} = \dfrac{c}{Hr}\dfrac{\partial}{\partial r}(\Phi^0 + \Phi) \end{cases} \quad (1)$$

The static electric field potential $\Phi^0 = U \cdot \ln(r/r_1)/\ln(r_2/r_1)$ gives $E_r = \mathrm{grad}\Phi^0$, $\partial\Phi^0/\partial\varphi = 0$, therefore, $E_\varphi(r)=0$. Here $U$ is the magnetron cathode voltage; $r_1$ and $r_2$ are the magnetron cathode and anode radii, respectively.

We consider a slow RF wave type $\exp[-i(n\varphi+\omega t)]$ excited at the frequency $\omega$ and rotating in the space of interaction with the angular phase velocity $\Omega = \omega/n$, [6]. The wave number $n=N/2$ sets the $\pi$-mode azimuthal pe-riodicity. The azimuthal velocity of the wave coincides with the azimuthal drift velocity of the center of the Larmor orbit located on the "synchronous" radius, $r_S$, [7], $r_s = \sqrt{-ncU/(\omega H \ln(r_2/r_1))}$, ($H<0$ is assumed).

* Supported by Fermi Research Alliance, LLC under Contract No. De-AC02- 07CH11359 with the United States DOE in collaboration with Muons, Inc.
[#]e-mail: gkazakevitch@yahoo.com; grigory@muonsinc.com

In magnetrons $r_L << 2\pi c/n\Omega$, where the right part of the inequality is the length of the synchronous wave. Therefore, one considers the interaction of the synchronous wave with an electron rotating along the Larmor orbit as an interaction of the wave with the point charge located in the center of the orbit. The electric field of the synchronous wave has radial and azimuthal components.

In a conventional magnetron ($\omega << \pi c/r_1$, $\pi c/r_2$), therefore the quasi-static approximation can be used to describe the rotating synchronous wave in the magnetron space of interaction. The scalar potential $\Phi$, satisfying the Laplace equation for the rotating wave is presented as in ref. [4]:

$$\Phi = \sum_{k=-\infty}^{\infty} \frac{\widetilde{E}_k(r_1) \cdot r_1}{2k} \left[ \left(\frac{r}{r_1}\right)^k - \left(\frac{r_1}{r}\right)^k \right] \sin(k\varphi + \omega t), \quad (2)$$

where $\widetilde{E}_k(r_1)$ is the amplitude of the $k$-th harmonic of the radial RF electric field at $r = r_1$. The form of the potential is chosen so that the azimuthal electric field vanishes at the cathode. The coefficients $\tilde{E}_k$ are determined to have zero azimuthal electric field at the anode everywhere except the coupling slits of the cavities. The term in the sum of Eq. (2) with $k=n$ has a resonant interaction with the azimuthal motion of the Larmor orbit. We consider only this term. It follows from Eqs. (1) that without the synchronous wave $\dot{r} = 0$ and drift of charges towards the anode, i.e., the magnetron generation, is impossible.

In the coordinate frame rotating with the synchronous wave for $\varphi_S = \varphi + t \cdot \omega/n$ and for an effective potential $\Phi_S$:

$$\Phi_S = U \frac{\ln(r/r_1)}{\ln(r_2/r_1)} + \frac{\omega H}{2nc} r^2 + \frac{\widetilde{E}_n r_1}{2n}\left[\left(\frac{r}{r_1}\right)^n - \left(\frac{r_1}{r}\right)^n\right] \sin(n\varphi_S)$$

one obtains the system of drift equations, [7]:

$$\begin{cases} \dot{r} = -\dfrac{c}{Hr} \dfrac{\partial}{\partial \varphi_S} \Phi_S, \\ \dot{\varphi}_S = \dfrac{c}{Hr} \dfrac{\partial}{\partial r} \Phi_S \end{cases} \quad (3)$$

Substituting the potential $\Phi_S$ into Eqs. (3) and denoting

$$\phi_0(r) = \ln\frac{r}{r_1} - \frac{1}{2}\left(\frac{r}{r_S}\right)^2, \quad \phi_1(r) = \frac{1}{2n}\left[\left(\frac{r}{r_1}\right)^n - \left(\frac{r_1}{r}\right)^n\right],$$

one can obtain the system of drift equations, [7, 4] in the frame of the synchronous wave expressed via the ratio, $\varepsilon$, of the radial field of the resonant harmonic of the synchronous wave, to the static electric field taken on the cathode.

Here: $\varepsilon = \widetilde{E}_n(r_1)/E_r(r_1) = \widetilde{E}_n(r_1) \cdot r_1 \ln(r_2/r_1)/U$.

$$\begin{cases} \dot{r} = \omega \dfrac{r_S^2}{r} \varepsilon \phi_1(r) \cos(n\varphi_S) \\ n\dot{\varphi}_S = -\omega \dfrac{r_S^2}{r}\left(\dfrac{d\phi_0}{dr} + \varepsilon \dfrac{d\phi_1}{dr} \sin(n\varphi_S)\right) \end{cases} \quad (4)$$

The Eqs. (4) characterize the resonant interaction of the charge in the center of the Larmor orbit with the synchronous wave.

The top equation in Eqs. (4) describes the radial velocity of the moving charge. The drift of the charge towards the anode is possible at $-\pi/2 < n\varphi_S < \pi/2$ with a period of $2\pi$, i.e., only in "spokes". The condition $\varepsilon \geq 1$ does not allow operation of the magnetron.

The second equation describes the azimuthal velocity of the drifting charge in the frame of the synchronous wave. The second term in the parentheses causes phase grouping of the charge by the resonant RF field via the potential $\phi_1$.

The first term of the equation describes a radially-dependent azimuthal drift of the charge resulting from the rotating frame with azimuthal angular velocity $-\omega/n$. This term at $r>r_S$ and low $\varepsilon$ causes the movement of the charge from the phase interval $\pm\pi/2$ allowed for "spokes" [4].

The Eqs. (4) were integrated for a typical model of a commercial magnetron described in ref. [4] with $N=8$, $r_1$ =5 mm, $r_2/r_1$ =1.5, $r_S/r_1$ =1.2. We obtained the charge trajectories at $r \geq r_1+r_L$ for various magnitudes $\varepsilon$ of the RF field in the synchronous wave and at the time interval ($\tau$) of the drift from 2 to 10 cyclotron periods allowing coherent contribution to the synchronous wave, [4]. The calculated azimuthal boundaries of the charge drifting in a "spoke" at various $\varepsilon$ are plotted in Fig. 1A. Normalized by $\pi$ the phase interval at $(r_1+r_L)$ determines part of the charge contributing to the synchronous wave. Fig. 1B shows trajectories of the charge in a "spoke" at $\varepsilon$ =0.3.

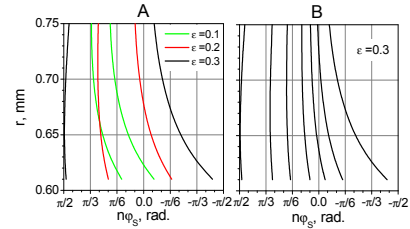

Figure 1: Phase grouping of the charge drifting towards the magnetron anode in the considered magnetron model.

The resonant interaction in magnetrons provides the energy exchange between the synchronous wave and the drifting charges grouped in "spokes". In the frame of the synchronous wave the RF azimuthal electric field in a "spoke" can be considered as stationary. The electric field strongly coupled with the resonant mode of the magnetron oscillation acts on the charge drifting in the "spoke". This causes the resonant energy exchange between the synchronous wave and the charge. If the azimuthal velocity of the drifting charge is greater than the azimuthal velocity of the synchronous wave, the charge being decelerated induces oscillation of the resonant mode in the magnetron RF system [8], and contributes it to the synchronous wave. Otherwise, the electric field of the wave accelerates the charge increasing its azimuthal drift velocity. This reduces the wave energy and the amplitude, respectively. Thus, the increase or decrease of the self-consistent electric field of the synchronous wave can be determined by the difference in the azimuthal velocities of the drifting charge and the synchronous wave,

$\Delta v_{AZ}(n\varphi_S, r)$, [9], Fig. 2. The necessary and sufficient conditions for stable generation of magnetrons driven by a resonant injected signal can be estimated by requiring positive values of the integral of $\Delta v_{AZ}(n\varphi_S, r)$ over the entire phase interval admissible for the "spoke". Fig. 2 shows that $\varepsilon \approx 0.3$ corresponding to the locking power of -10 dB provides coherent generation of the tube, while at $\varepsilon \leq 0.2$ the stable generation is impossible. The ratio of the synchronous wave radial field to the static electric field on the magnetron cathode of about 0.3 provides coherent generation of the tube even below the Hartree voltage [4].

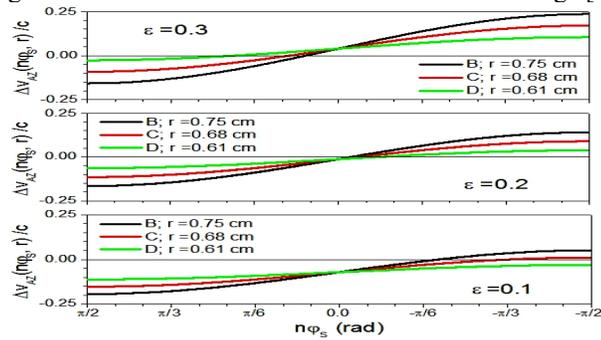

Figure 2: The difference of the azimuthal velocities of the drifting charge and the synchronous wave in units of $c$ at various $\varepsilon$ and $n\varphi_S$.

## EXPERIMENTAL VERIFICATION OF THE ANALYTICAL MODEL

Precise stability of the carrier frequency when the magnetron operates below and above the Hartree voltage at various powers of the resonant injected signal, $P_{Lock}$, is shown in Fig. 3 [4].

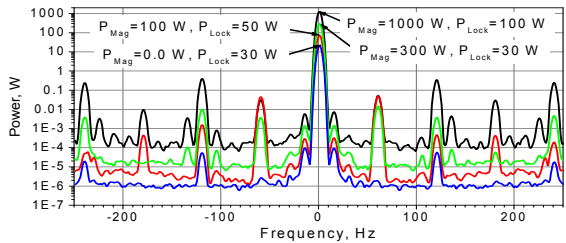

Figure 3: Offset of the carrier frequency at various power levels of the magnetron, and the locking signal. The graph $P_{Mag}$ =0.0 W, $P_{Lock}$ =30 W was measured with the magnetron high voltage turned OFF. Graphs at $P_{Mag}$=100 W and 300 W were measured below the Hartree voltage.

A highest efficiency and wide range of power control was verified in experiments [4], Fig. 4.

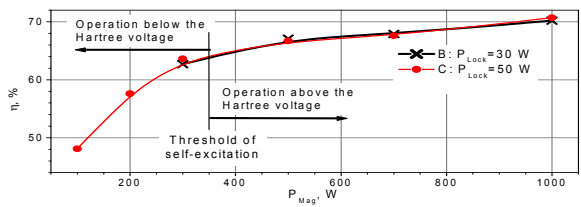

Figure 4: Efficiency of the 2.45 GHz, 1.2 kW magnetron operating below and above the Hartree voltage.

Fig. 5, [4], demonstrates low magnetron noise at $\varepsilon \approx 0.3$.

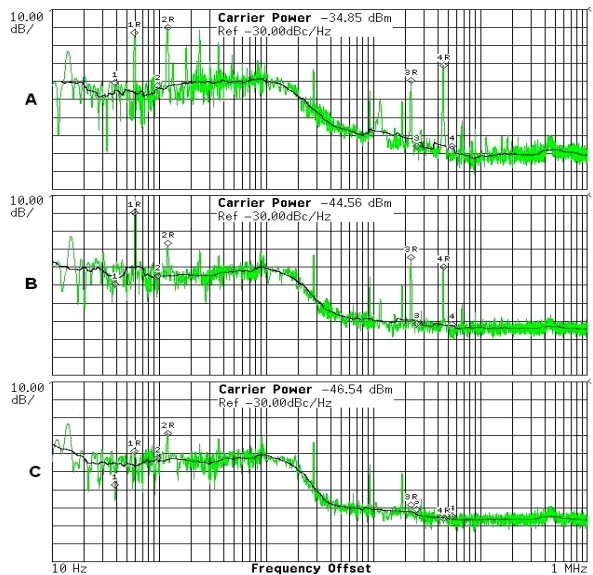

Figure 5: Spectral power density of noise of the magnetron driven by the resonant injected signal at $\varepsilon \approx 0.3$ when operated above and below the Hartree voltage. Trace A: $P_{Mag}$=1 kW, trace B: $P_{Mag}$=0.1 kW, trace C: the noise spectral density of the resonant injected signal of 0.1 kW; the magnetron HV is OFF.

## SUMMARY

The analytical model considers the resonant interaction of a magnetron's synchronous wave with drifting charges grouped in phase. It is the basis of operation of the magnetrons. The model allows for optimizing the injected resonant signal for operation of the tube with high stability, high efficiency and low noise in accordance with requirements of superconducting accelerators. The experiments performed demonstrated that the behaviour of magnetrons driven by the resonant injected signal and fed below and above the Hartree voltage is well explained and substantiated by the analytical model described.

## REFERENCES

[1] Z. Conway, and M. Liepe, report TU5PFP043, *in Proceed. of PAC09,* 1-3, 2009.
[2] G. Kazakevich, et al., NIM A 760 (2014) 19–27.
[3] B. Chase, et. al.,JINST, 10, P03007, 2015.
[4] G. Kazakevich, et al., NIM A 839 (2016) 43-51.
[5] H.L. Thal and R.G. Lock, IEEE Trans. MTT, V. 13, 1965.
[6] P.L. Kapitza, "High Power Electronics", Sov. Phys. Uspekhi, V 5, # 5, 777-826, 1963.
[7] L.A. Vainstein and V.A. Solntsev, *in* "Lectures on microwave electronics", Moscow, Sov. Radio, 1973, (In Russian).
[8] D. H. Whittum, Frontiers of Accelerator Technology, U.S.-CERN-Japan International School, Japan, edited by S. I. Kurokawa, M. Month, and S. Turner (World Scientific, Singapore, London, 1996), pp. 1–135.
[9] G. Kazakevich et al., arXive:1709.04526.